\documentclass[final,1p,times,onecolumn]{elsarticle}

\usepackage{soul}
\usepackage{amssymb}
\usepackage{lipsum}
\usepackage{booktabs}
\usepackage{dirtree}
\usepackage{doi}
\usepackage{amsmath}
\usepackage[numbers]{natbib}

\journal{Chemical Data Collection}

\begin{document}

\begin{frontmatter}

\title{DFT GGA based datasets for H$_2$O potential energy surfaces, permanent moment and polarizability tensors}

\author[first,label1]{Anoop Ajaya Kumar Nair \corref{cor1}}
\ead{mailanoopanair@gmail.com}

\author[second,label1]{Elvar Örn J\'onsson \corref{cor1}}
\ead{elvarorn@hi.is}

\cortext[cor1]{Corresponding author.}

\affiliation[label1]{organization={Science Institute and Faculty of Physical Sciences, University of Iceland},
            addressline={S\ae mundargata},
            city={Reykjav{\'i}k},
            postcode={107},
            country={Iceland}}

\begin{abstract}
We present in this article data on the computed potential energy surface, permanent moment surface (from dipole to hexadecapole) and polarizability surface (from dipole-dipole to quadrupole-quadrupole) for the H$_2$O molecule, calculated at the level of the generalized gradient approximation (GGA) in Kohn-Sham density functional theory (KS-DFT) in the grid-based projector augmented wave code GPAW, using the charge perturbation code C-pol. The GGA-based functionals are PBE, RPBE and BLYP. The monomer surfaces are composed of systematic, symmetry-reduced deformations of the H$_2$O monomer and comprise 10{,}000 configurations per functional, with traceless Cartesian tensor moment values reported as data strings in the \emph{xyz} file format. In addition, we provide cluster datasets for n-H$_2$O aggregates (n = 2-6), sampled from liquid-like configurations and computed with the same KS-DFT settings, where each extended \emph{xyz} file encodes the total interaction energy of the cluster in the comment line. Finally, ab initio molecular dynamics (AIMD) trajectories of periodic liquid water boxes, containing between 25 to 150 H$_2$O molecules at ambient conditions, are included, furnishing time series of atomic configurations and energies that are consistent with the monomer and cluster data and suitable for validating classical and machine-learned water models.
\end{abstract}


\end{frontmatter}

\newpage
\section*{Specification Table}
\begin{table}[th!]
\centering
\begin{tabular}{p{4cm} p{9cm}}
\toprule
\textbf{Subject} & Physical \& Computational Chemistry\\
\midrule
\textbf{Specific subject area} & DFT calculations of energy and multipoles of H$_2$O structures (monomers). Energy calculations of n-H$_2$O clusters [where n=2 to 6] and molecular dynamics trajectories.\\
\midrule
\textbf{Type of data} & {\it xyz} files which include energy and multipole property as parse-able information string \\
\midrule
\textbf{How data were acquired} & Electronic structure calculations of ground and perturbed states using GPAW \citep{GPAW} with \texttt{C-Pol} \citep{Cpol} implementation.\\
\midrule
\textbf{Data format} & Extended \textit{xyz} files in which each configuration's comment line encodes the computed properties as key-value pairs, ordered as: traceless Cartesian multipole moments (dipole through hexadecapole) followed by traceless Cartesian polarizability tensors (dipole-dipole, dipole-quadrupole, and quadrupole-quadrupole). \\
\midrule
\textbf{Parameters for data collection} & GPAW calculations were performed with the GGA functionals PBE \citep{PBE}, RPBE \citep{RPBE}, and BLYP \citep{BinBLYP,LYPinBLYP}, with grid spacing $h=0.18\ \AA$, in finite-difference mode. The self-consistent-field cycle employed strict convergence criteria on the multipole moments, Kohn-Sham eigenstates, and electron density (``\texttt{qpoles}'' and ``\texttt{dpoles}'' set to $10^{-6}$ [a.u.], and ``\texttt{eigenstates}'' and ``\texttt{density}'' set to $10^{-8}$ [a.u.]). \\
\midrule
\textbf{Description of data collection} & Monomer structure data (energy and multipoles) was obtained using routines available in C-Pol (implemented in GPAW) \citep{GPAW, Cpol}. The cluster data and H$_2$O Box (Molecular dynamics)  simulations were performed using GPAW.\\
\midrule
\textbf{Data source location} & VR-3, University of Iceland, Reykjav{\'i}k, Iceland\\
\midrule
\textbf{Data accessibility} & Available in Zenodo repository: \href{https://doi.org/10.5281/zenodo.21053088}{{10.5281/zenodo.21053088}} . Code is available: gitlab repo \citep{h2odataset} \\
\midrule
\textbf{Related research article} & C-Pol \citep{Cpol} - generation of permanent and polarizable moments. \\
\bottomrule
\end{tabular}
\end{table}

\clearpage

\section{Rationale}

Machine-learned interatomic potentials (MLIPs) typically rely on local, fixed-cutoff atomic decompositions, which limits their ability to capture long-range electrostatics and polarization, and often results in unphysical dielectric response and poor transferability across phases. Hybrid architectures that couple an explicit long-range polarization model with a short-range machine-learned correction term such as SCME-MACE, which combines the Single-Center Multipole Expansion (SCME) with an equivariant neural network (MACE), offer one route to addressing this, but require reference data beyond total energies alone: geometry-resolved permanent multipole moments and polarizability tensors, together with energies for hydrogen-bonded clusters and condensed-phase configurations.
This dataset provides that data in two parts. First, a systematic sampling of the H$_2$O monomer's internal coordinate space with computed permanent multipole moments (dipole through hexadecapole) and polarizability tensors (dipole-dipole through quadrupole-quadrupole), suitable for fitting geometry-dependent electrostatic models. Second, energies for water clusters (dimers through hexamers) and ab initio molecular dynamics (AIMD) trajectories of liquid water, capturing cooperative hydrogen-bonding effects relevant to training or fine-tuning short-range correction terms.
Beyond its original motivation, the cluster and AIMD data also serve as a standardized benchmark for existing foundation MLIPs (e.g., MACE, NequIP, MACE-LES) that handle long-range interactions with differing approximations, allowing direct comparison against consistent PBE-level reference energetics. The inclusion of three GGA functionals (PBE, RPBE, BLYP) further allows assessment of how sensitive fitted or fine-tuned models are to the choice of exchange-correlation treatment.

\section*{Value of the data}

\begin{enumerate}
 \item Computational data usually only includes energy, and possibly forces, as function of geometry. This dataset includes other physical properties: permanent moments and polarizabilities. Energy data for water clusters spanning dimers to hexamers captures the full range of 
short- and long-range electrostatic interactions and cooperative polarization effects, 
providing a rigorous reference for the development and validation of accurate liquid-phase 
water models.
 \item Data can be used to map energy, permanent and polarizable moment surfaces using e.g. classical polynomial expressions such as spectral functions \citep{Loboda2016}, and complement classical flexible polarizable potentials like SCME/f \citep{scmef}. 
 \item The data can be used to enrich the training of machine-learned interatomic potentials (MLIPs) like MACE-POLAR \citep{mace-polar} or MACE-MDP \citep{mace-mdp}. The molecular dynamics trajectories and cluster energies can be used to fine-tune general-purpose MLIPs such as MACE \citep{Batatia2022mace,Batatia2022Design}, NequIP \citep{NequIP}, and UMA \citep{UMA} improving their 
accuracy in simulation environments where H$_2$O acts as a solvent.
\end{enumerate}

\section{Description of the data}

\begin{figure}[!ht]
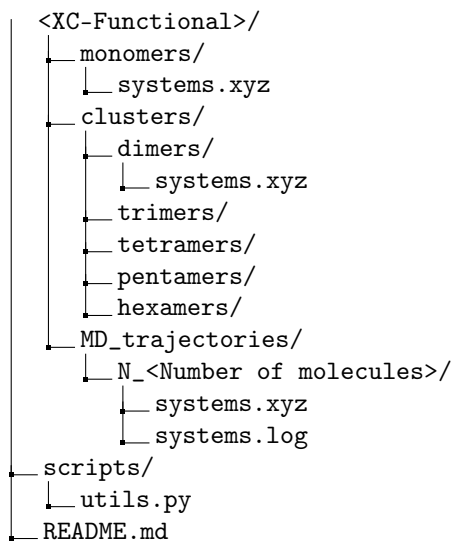

  \dirtree{%
  .1 <XC-Functional>/.
  .2 monomers/.
  .3 systems.xyz.
  .2 clusters/.
  .3 dimers/.
  .4 systems.xyz.
  .3 trimers/.
  .3 tetramers/.
  .3 pentamers/.
  .3 hexamers/.
  .2 MD\_trajectories/.
  .3 N\_<Number of molecules>/.
  .4 systems.xyz.
  .4 systems.log.
  .1 scripts/.
  .2 utils.py.
  .1 README.md.
  }
  \caption{Directory structure of the data repository.}
  \label{fig:project-tree}
\end{figure}

The dataset is organised by exchange-correlation functional, with separate subdirectories for isolated monomers and for clusters of increasing size, as illustrated in Figure.\ref{fig:project-tree}.\ Each \texttt{monomers/} directory contains a single \texttt{systems.xyz} file collecting all monomer geometries and properties (as mentioned in the \textbf{monomer data} section), while the \texttt{clusters/} directory is further divided into \texttt{dimers/}, \texttt{trimers/}, \texttt{tetramers/}, \texttt{pentamers/} and \texttt{hexamers/}, each storing the corresponding structures (in single \texttt{system.xyz} files). The cluster data only has the geometries (in XYZ format) and the corresponding energies provided in the comment line as \texttt{"energy = <value>"} pair. Molecular dynamics trajectories are stored in \texttt{MD\_trajectories/ (H$_2$O Box)}, where each subdirectory \texttt{N\_\textless Number of molecules\textgreater/} contains a \texttt{system.xyz} file for that system size (indicated by the number of molecules). The number of data points in the directories for each XC-functional is provided in Table \ref{tab:data-counts},  and auxiliary analysis tools are placed in \texttt{scripts/} (for example, \texttt{utils.py}). \texttt{utils.py} provides two helper parsers: one that converts extended XYZ cluster files into ASE \texttt{Atoms} objects with attached single-point energies, and another that reads monomer XYZ files and returns per-frame dictionaries containing geometry, scalar metadata (ID, geometry parameters, energy), and embedded multipole moments and polarizability tensors for each water monomer configuration. A brief overview of the dataset is provided in the top-level \texttt{README.md}.

\begin{table}[!ht]
\centering
\caption{Overview of data types and number of instances for each functional. Each functional-specific dataset contains 10,000 monomer, 57,000 cluster, and 8,000 MD-box configurations, for 3x that total across PBE/RPBE/BLYP.
}
\label{tab:data-counts}
\begin{tabular}{p{0.3\textwidth}  p{0.3\textwidth} p{0.3\textwidth}}
\toprule
\textbf{Data type/ Configuration} & \textbf{No. of datapoints (approx.)} & \textbf{Breakdown} \\
\midrule
Monomers  & 10\,000 & N/A \\
Dimers    & 30\,000  & N/A\\
Trimers   & 10\,000  & N/A\\
Tetramers & 3\,000  & N/A\\
Pentamers & 6\,000  & N/A\\
Hexamers  & 8\,000  & N/A\\
H$_2$O Box    & 8\,000  & Boxes containing 25 to 150 water molecules\\
\bottomrule
\end{tabular}
\end{table}

\noindent \textbf{Monomer data: } In each \texttt{system.xyz} file, the comment line associated with every configuration contains the computed properties in the format \texttt{key="value string"}. The ordering of the quantities within this string follows the convention specified in Table~\ref{tab:monomer-properties}.

\begin{table*}[!h] 
\centering
\begin{tabular}{llll}
\hline
\textbf{Key} & \textbf{Size} & \textbf{Meaning} & \textbf{Ordering of components} \\
\hline
\texttt{energy} &
Scalar &
Total monomer energy &
$E$ stored as a float \\[4pt]

\texttt{dp} &
3 &
Dipole moment vector &
$\mu_x$, $\mu_y$, $\mu_z$ \\[4pt]

\texttt{qp} &
6 &
Quadrupole tensor (rank 2, symmetric) &
$Q_{xx}$, $Q_{xy}$, $Q_{xz}$, $Q_{yy}$, $Q_{yz}$, $Q_{zz}$ \\[4pt]

\texttt{op} &
10 &
Octupole tensor (rank 3, symmetric) &
$O_{xxx}$, $O_{xxy}$, $O_{xxz}$,\\
& & &
$O_{xyy}$, $O_{xyz}$, $O_{xzz}$,\\ 
& & &
$O_{yyy}$, $O_{yyz}$, $O_{yzz}$,\\
& & &
$O_{zzz}$ \\[4pt]

\texttt{hp} &
15 &
Hexadecapole tensor (rank 4, symmetric) &
$H_{xxxx}$, $H_{xxxy}$, $H_{xxxz}$,\\
& & &
$H_{xxyy}$, $H_{xxyz}$, $H_{xxzz}$,\\
& & &
$H_{xyyy}$, $H_{xyyz}$, $H_{xyzz}$,\\
& & &
$H_{xzzz}$, $H_{yyyy}$, $H_{yyyz}$,\\
& & &
$H_{yyzz}$, $H_{yzzz}$, $H_{zzzz}$ \\[4pt]

\texttt{dd} &
6 &
Dipole-dipole tensor (rank 2) &
$\alpha_{xx}$, $\alpha_{xy}$, $\alpha_{xz}$,\\
& & &
$\alpha_{yy}$,  $\alpha_{yz}$, $\alpha_{zz}$ \\[4pt]

\texttt{dq} &
18 &
Dipole--quadrupole tensor (rank 3) &
$A_{x,xx}$, $A_{x,xy}$, $A_{x,xz}$,\\
& & &
$A_{x,yy}$, $A_{x,yz}$, $A_{x,zz}$,\\
& & & 
$A_{y,xx}$, $A_{y,xy}$, $A_{y,xz}$,\\
& & &
$A_{y,yy}$, $A_{y,yz}$, $A_{y,zz}$,\\
& & &
$A_{z,xx}$, $A_{z,xy}$, $A_{z,xz}$, \\
& & &
$A_{z,yy}$, $A_{z,yz}$, $A_{z,zz}$ \\[4pt]

\texttt{qq} &
21 &
Quadrupole--quadrupole tensor (rank 4) &
Subset of independent components \\
& & &
$C_{xx,xx}$, $C_{xx,xy}$, $C_{xx,xz}$,\\
& & &
$C_{xx,yy}$, $C_{xx,yz}$, $C_{xx,zz}$,\\ 
& & &
$C_{xy,xy}$, $C_{xy,xz}$, $C_{xy,yy}$,\\ 
& & &
$C_{xy,yz}$, $C_{xy,zz}$, $C_{xz,xz}$,\\
& & &
$C_{xz,yy}$, $C_{xz,yz}$, $C_{xz,zz}$,\\
& & &
$C_{yy,yy}$, $C_{yy,yz}$, $C_{yy,zz}$,\\ 
& & &
$C_{yz,yz}$, $C_{yz,zz}$, $C_{zz,zz}$\\
\hline
\end{tabular}
\caption{Monomer properties (energy, moments and polarizabilities in traceless form) stored in the \texttt{Atoms.info} dictionary and written to extxyz. All multipole moments and polarizability tensors are reported in the Cartesian local frame defined by the GPAW simulation cell, which is an axis-aligned cubic box. No rotation to a molecule-fixed frame is applied prior to storage; the molecular geometry (atomic positions) stored in the same XYZ frame provides the full information needed to reconstruct any molecule-fixed transformation.}
\label{tab:monomer-properties}
\end{table*}

\section{Computational details}

Distances are reported in \r{a}ngstrom (1 \AA~= 0.1 nm = 10$^{-10}$ m) throughout, consistent with common practice in the electronic-structure literature.

\subsection{Monomer calculations}
Monomer calculations were performed with \texttt{GPAW} using the exchange-correlation functional of interest (PBE, RPBE and BLYP) and a real-space grid spacing of $h = 0.18$~\AA. The self-consistent-field cycle employed strict convergence criteria on the multipole moments, Kohn-Sham eigenstates, and electron density (``\texttt{qpoles}'' and ``\texttt{dpoles}'' set to $10^{-6}$ [a.u.], and ``\texttt{eigenstates}'' and ``\texttt{density}'' set to $10^{-8}$ [a.u.]). The Kohn-Sham equations were solved with the RMM-DIIS eigensolver, limiting the maximum number of eigensolver iterations per SCF step to \texttt{niter=5}. A vacuum of 7~\AA\ was used to prevent periodic image interactions. The moments and polarizabilites were calculated using the \texttt{C-Pol} implementation in \texttt{GPAW}. 
The monomer configurations were generated using systematic, symmetry-reduced deformations of H$_2$O, with O-H bond lengths sampled from 0.55 to 1.55~\AA\ in 0.05~\AA\ increments and H-O-H angles from $55^\circ$ to $175^\circ$ in $5^\circ$ increments. After computing the multipoles, symmetry operations were applied to augment the dataset and achieve a more comprehensive sampling of the configuration space. For the \( \mathrm{H_2O} \) molecule with \( C_{2v} \) or \( C_s \) symmetry, a reflection in the plane of the molecular bisector axis was used. The multipole moments are stored as traceless Cartesian tensors
following the convention implemented in C-Pol \citep{Cpol}, which follows
the Buckingham--Stone convention as described in \textit{The Theory of
Intermolecular Forces} \citep{Stone2013}. Specifically, the quadrupole tensor
\(Q_{\alpha\beta}\) is defined as
\begin{equation}
    Q_{\alpha\beta} = \frac{1}{2}\sum_{i} q_{i}
    \left( 3r_{i\alpha}\,r_{i\beta} - r_{i}^{2}\,\delta_{\alpha\beta} \right),
\end{equation}
and the analogous traceless forms are used for higher-rank tensors \citep{Cpol}. 

\subsection{Cluster calculations}

Cluster calculations (for trimer to hexamer sampled from MD trajectory) were performed with \texttt{GPAW} using the same parameters as used for the monomer systems. All clusters were embedded in a cubic simulation cell of length 25~\AA, and the molecular center of mass was shifted to the box center to minimise interactions with periodic images. Energies were evaluated for each geometry in batches and written to disk for subsequent analysis. Multipole moments and polarizabilities were not evaluated for the cluster
configurations, as monomer values have been shown to be transferable and
sufficient for parameterizing classical flexible polarizable force fields of
the SCME/\textit{f} type \citep{scmef}.  Atomic forces were
not evaluated for any configuration in this dataset, as most machine learning
interatomic potentials obtain forces analytically as derivatives of the energy,
and only energy evaluations were required for the purposes of our prior work
\citep{scmef}. 

\subsection{H$_2$O Box/ MD simulations}

Ab initio molecular dynamics (AIMD) simulations of liquid water boxes were carried out using GPAW in real-space mode. A representative system consisted of a cubic periodic cell of edge length 20~\AA\ containing 25 to 150 H$_2$O molecules at 300~K. Initial velocities were assigned from a Maxwell-Boltzmann distribution at the target temperature, followed by removal of net linear and angular momentum. The equations of motion were propagated with a Langevin thermostat using a time step of 0.5~fs and a friction coefficient of 0.01~fs$^{-1}$, for a total of 5000 MD steps (2.5~ps of simulation time), with trajectories and basic thermodynamic quantities recorded at regular intervals. Equivalent simulations were performed for each of the three GGA exchange--correlation functionals considered in this work (PBE, RPBE, and BLYP).

\section{Value and Validation}

\begin{figure}
    \centering
    \includegraphics[width=1\linewidth]{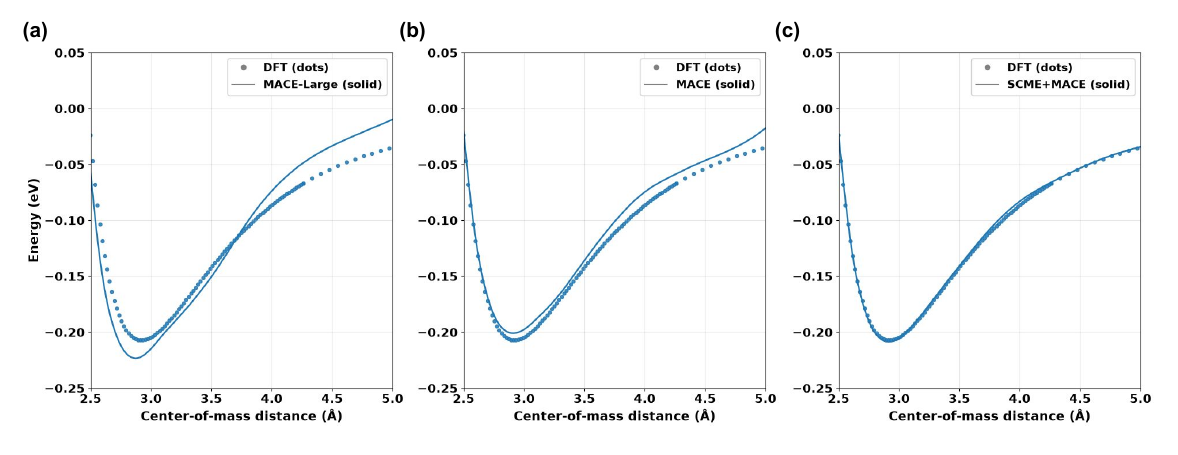}
    \caption{Interaction energy versus center-of-mass distance of the $\text{H}_2\text{O}$ dimer predicted using: \textbf{(a)} untuned MACE-MP-0 (large variant), \textbf{(b)} MACE-MP-0 (large variant) fine-tuned on energies from the $\text{H}_2\text{O}$ PBE dataset, and \textbf{(c)} fine-tuned MACE-MP-0 (large variant) augmented with an electrostatic model trained on multipoles (moments and polarizabilities) from the $\text{H}_2\text{O}$ PBE dataset. The dots in each plot represent the DFT-based reference energy data (PBE, GPAW).}
    \label{fig:dimerPBEplots}
\end{figure}

The permanent moment values reported for the monomer surfaces follow directly from the underlying GPAW electron density and are subject to the same convergence criteria used throughout the calculations (grid spacing, vacuum size, SCF tolerances), while the polarizability tensors (dipole-dipole, dipole-quadrupole, and quadrupole-quadrupole) were computed using the C-Pol point-charge perturbation scheme, whose accuracy has been independently validated in the companion methods article \citep{Cpol}. There, polarizability tensors computed with C-Pol in GPAW were benchmarked against an energy-based finite-field reference (Gaussian 16, extrapolated to the complete basis set limit) across one representative molecule from each of the 19 commonly occurring point groups, reproducing all symmetry-mandated tensor patterns and agreeing with the reference values to within 3\% for components that contribute meaningfully to the electrostatic potential. The dipole-dipole polarizability was further validated against the analytic coupled-perturbed Hartree-Fock method, with agreement to within 0.02 a.u., and the combined moment/polarizability description was shown to reproduce the full PBE based QM electrostatic potential of H$_2$O and CO$_2$ dimers to within RMS errors on the order of 10$^{-4}$–10$^{-5}$ a.u. Polarizability tensors were computed only for the isolated monomer configurations; cluster (dimer–hexamer) and AIMD trajectory data in this dataset provide energies only, on the basis that monomer-derived multipole and polarizability values have previously been shown to be transferable for parameterizing flexible, geometry-dependent polarizable water force fields such as SCME/f. This is particularly evident (in Figure~\ref{fig:dimerPBEplots}) in the dimer interaction energy plots (interaction energy as a function of distance between center of mass of the monomers) predicted using the different iterations of our machine learning potential framework.  As shown in Figure~\ref{fig:dimerPBEplots}a, the untuned baseline MACE-MP-0 (large variant; \texttt{MACE-Large}) exhibits a noticeable deviation from the target reference data, overestimating the binding depth near the minimum and crossing the reference curve around a center-of-mass distance of $3.6$~\AA. Upon fine-tuning the model directly on the energies of the $\text{H}_2\text{O}$ PBE dataset (Figure~\ref{fig:dimerPBEplots}b), the prediction curve (\texttt{MACE}) alignment improves dramatically across the entire spatial range. The fine-tuned MACE model successfully captures the location of the energy minimum near $2.9$~\AA, though minor deviations remain at longer separation distances ($>4.0$~\AA) where long-range tail behaviors dictate the curve.  The most robust agreement is achieved when physics-informed constraints are explicitly introduced, as illustrated in Figure~\ref{fig:dimerPBEplots}c. By augmenting the fine-tuned model with an electrostatic architecture (\texttt{SCME+MACE}) trained on PBE-derived multipole moments and polarizabilities, the physical long-range tail is smoothly resolved. This hybrid formulation yields an excellent overlay with the reference DFT data points across the short-range repulsive wall, the equilibrium binding well, and the long-range asymptotic decay region. This is a clear example of the utility of the dataset associated with the publication.

\section*{Acknowledgements}
This work was supported by the Icelandic Research Fund, grant no.\ 2410644 and the Eimskip fund, grant no. HEI2023-93153. Computer resources, data storage, and user support were provided by the 
Icelandic Research e-Infrastructure (IREI), funded by the Icelandic Infrastructure Fund.

\appendix

\bibliographystyle{unsrtnat} 
\bibliography{example}

@misc{mace-mdp,
author = {Gönnheimer, Nils and Reuter, Karsten and Kapil, Venkat and Margraf, Johannes},
year = {2026},
month = {04},
pages = {},
title = {MACE-MDP: A General Dipole and Polarizability Model for Organic Molecules and Materials},
doi = {10.26434/chemrxiv.15000716/v2},
url = {https://chemrxiv.org/doi/full/10.26434/chemrxiv.15000716/v2}
}

@misc{mace-polar,
      title={MACE-POLAR-1: A Polarisable Electrostatic Foundation Model for Molecular Chemistry}, 
      author={Ilyes Batatia and William J. Baldwin and Domantas Kuryla and Joseph Hart and Elliott Kasoar and Alin M. Elena and Harry Moore and Mikołaj J. Gawkowski and Benjamin X. Shi and Venkat Kapil and Panagiotis Kourtis and Ioan-Bogdan Magdău and Gábor Csányi},
      year={2026},
      eprint={2602.19411},
      archivePrefix={arXiv},
      primaryClass={physics.chem-ph},
      url={https://arxiv.org/abs/2602.19411}, 
}

@article{scmef,
title = "Transferable Potential Function for Flexible H2O Molecules Based on the Single-Center Multipole Expansion",
author = "J{\'o}nsson, Elvar {\"O}rn and Soroush Rasti and Marta Galynska and J{\"o}rg Meyer and Hannes J{\'o}nsson",
year = "2022",
month = dec,
day = "13",
doi = "10.1021/acs.jctc.2c00598",
language = "English / enska",
volume = "18",
pages = "7528--7543",
journal = "Journal of Chemical Theory and Computation",
issn = "1549-9618",
publisher = "American Chemical Society",
number = "12",
}

@article{Loboda2016,
author = {Loboda, Oleksandr and Ingrosso, Francesca and Ruiz-L{\'o}pez, Manuel F. and Reis, Heribert and Millot, Claude},
title = {Dipole and quadrupole polarizabilities of the water molecule as a function of geometry},
journal = {Journal of Computational Chemistry},
volume = {37},
number = {23},
pages = {2125-2132},
doi = {https://doi.org/10.1002/jcc.24431},
url = {https://onlinelibrary.wiley.com/doi/abs/10.1002/jcc.24431},
year = {2016}
}

@article{PBE,
  title = {Generalized Gradient Approximation Made Simple},
  author = {Perdew, John P. and Burke, Kieron and Ernzerhof, Matthias},
  journal = {Phys. Rev. Lett.},
  volume = {77},
  issue = {18},
  pages = {3865--3868},
  numpages = {0},
  year = {1996},
  month = {Oct},
  publisher = {American Physical Society},
  doi = {10.1103/PhysRevLett.77.3865},
  url = {https://link.aps.org/doi/10.1103/PhysRevLett.77.3865}
}

@article{RPBE,
  title = {Improved adsorption energetics within density-functional theory using revised Perdew-Burke-Ernzerhof functionals},
  author = {Hammer, B. and Hansen, L. B. and N\o{}rskov, J. K.},
  journal = {Phys. Rev. B},
  volume = {59},
  issue = {11},
  pages = {7413--7421},
  numpages = {0},
  year = {1999},
  month = {Mar},
  publisher = {American Physical Society},
  doi = {10.1103/PhysRevB.59.7413},
  url = {https://link.aps.org/doi/10.1103/PhysRevB.59.7413}
}

@article{BinBLYP,
  title = {Density-functional exchange-energy approximation with correct asymptotic behavior},
  author = {Becke, A. D.},
  journal = {Phys. Rev. A},
  volume = {38},
  issue = {6},
  pages = {3098--3100},
  numpages = {0},
  year = {1988},
  month = {Sep},
  publisher = {American Physical Society},
  doi = {10.1103/PhysRevA.38.3098},
  url = {https://link.aps.org/doi/10.1103/PhysRevA.38.3098}
}

@article{LYPinBLYP,
  title = {Development of the Colle-Salvetti correlation-energy formula into a functional of the electron density},
  author = {Lee, Chengteh and Yang, Weitao and Parr, Robert G.},
  journal = {Phys. Rev. B},
  volume = {37},
  issue = {2},
  pages = {785--789},
  numpages = {0},
  year = {1988},
  month = {Jan},
  publisher = {American Physical Society},
  doi = {10.1103/PhysRevB.37.785},
  url = {https://link.aps.org/doi/10.1103/PhysRevB.37.785}
}

@article{GPAW,
    author = {Mortensen, Jens Jørgen and Larsen, Ask Hjorth and Kuisma, Mikael and Ivanov, Aleksei V. and Taghizadeh, Alireza and Peterson, Andrew and Haldar, Anubhab and Dohn, Asmus Ougaard and Schäfer, Christian and Jónsson, Elvar Örn and Hermes, Eric D. and Nilsson, Fredrik Andreas and Kastlunger, Georg and Levi, Gianluca and Jónsson, Hannes and Häkkinen, Hannu and Fojt, Jakub and Kangsabanik, Jiban and Sødequist, Joachim and Lehtomäki, Jouko and Heske, Julian and Enkovaara, Jussi and Winther, Kirsten Trøstrup and Dulak, Marcin and Melander, Marko M. and Ovesen, Martin and Louhivuori, Martti and Walter, Michael and Gjerding, Morten and Lopez-Acevedo, Olga and Erhart, Paul and Warmbier, Robert and Würdemann, Rolf and Kaappa, Sami and Latini, Simone and Boland, Tara Maria and Bligaard, Thomas and Skovhus, Thorbjørn and Susi, Toma and Maxson, Tristan and Rossi, Tuomas and Chen, Xi and Schmerwitz, Yorick Leonard A. and Schiøtz, Jakob and Olsen, Thomas and Jacobsen, Karsten Wedel and Thygesen, Kristian Sommer},
    title = {GPAW: An open Python package for electronic structure calculations},
    journal = {The Journal of Chemical Physics},
    volume = {160},
    number = {9},
    pages = {092503},
    year = {2024},
    month = {03},
    doi = {10.1063/5.0182685},
    url = {https://doi.org/10.1063/5.0182685},
}

@misc{Cpol,
      title={C-Pol: Point charge perturbation scheme for mapping tensor moment surfaces}, 
      author={Anoop Ajaya Kumar Nair and Julian Beßner and Timo Jacob and Elvar Örn Jónsson},
      year={2026},
      eprint={2409.10184v2},
      archivePrefix={arXiv},
      primaryClass={physics.chem-ph},
      url={https://arxiv.org/abs/2409.10184v2}, 
}

@article{NequIP,
  author    = {Batzner, Simon and Musaelian, Albert and Sun, Lixin and Geiger, Mario and
               Mailoa, Jonathan P. and Kornbluth, Mordechai and Molinari, Nicola and
               Smidt, Tess E. and Kozinsky, Boris},
  title     = {E(3)-equivariant graph neural networks for data-efficient and accurate interatomic potentials},
  journal   = {Nature Communications},
  year      = {2022},
  volume    = {13},
  number    = {1},
  pages     = {2453},
  doi       = {10.1038/s41467-022-29939-5},
  url       = {https://doi.org/10.1038/s41467-022-29939-5},
}

@misc{UMA,
      title={UMA: A Family of Universal Models for Atoms}, 
      author={Brandon M. Wood and Misko Dzamba and Xiang Fu and Meng Gao and Muhammed Shuaibi and Luis Barroso-Luque and Kareem Abdelmaqsoud and Vahe Gharakhanyan and John R. Kitchin and Daniel S. Levine and Kyle Michel and Anuroop Sriram and Taco Cohen and Abhishek Das and Ammar Rizvi and Sushree Jagriti Sahoo and Zachary W. Ulissi and C. Lawrence Zitnick},
      year={2026},
      eprint={2506.23971},
      archivePrefix={arXiv},
      primaryClass={cs.LG},
      url={https://arxiv.org/abs/2506.23971}, 
}

@inproceedings{Batatia2022mace,
  title={{MACE}: Higher Order Equivariant Message Passing Neural Networks for Fast and Accurate Force Fields},
  author={Ilyes Batatia and David Peter Kovacs and Gregor N. C. Simm and Christoph Ortner and Gabor Csanyi},
  booktitle={Advances in Neural Information Processing Systems},
  editor={Alice H. Oh and Alekh Agarwal and Danielle Belgrave and Kyunghyun Cho},
  year={2022},
  url={https://openreview.net/forum?id=YPpSngE-ZU}
}

@misc{h2odataset,
  author       = {Nair, Anoop Ajaya Kumar and J{\'o}nsson, Elvar {\"O}rn},
  title        = {{DFT GGA-based dataset for H$_2$O potential energy,
                   permanent moment and polarizability surfaces}},
  year         = {2026},
  publisher    = {GitLab},
  journal      = {GitLab repository},
  howpublished = {\url{https://gitlab.com/Anoop_publications/h2o_dataset}},
  note         = {Accessed: 2026-06-30}
}

@book{Stone2013,
  author    = {Stone, Anthony J.},
  title     = {The Theory of Intermolecular Forces},
  edition   = {2nd},
  publisher = {Oxford University Press},
  address   = {Oxford},
  year      = {2013},
  doi       = {10.1093/acprof:oso/9780199672394.001.0001},
}


\end{document}